\documentclass[amsmath,amssymb,twocolumn,superscriptaddress]{revtex4-1}

\usepackage{graphicx}
\usepackage{dcolumn}
\usepackage{bm}
\usepackage[detect-all]{siunitx}
\usepackage{hyperref}
\usepackage{xr}
\usepackage{color}
\usepackage{mhchem}
\graphicspath{{./}}

\begin{document}                  

\title{Assessing molecular simulation for the analysis of lipid monolayer reflectometry}

\author{A.~R. McCluskey}
\email{a.r.mccluskey@bath.ac.uk}
\email{andrew.mccluskey@diamond.ac.uk}
\affiliation{Department of Chemistry, University of Bath, Claverton Down,
Bath, BA2 7AY, UK}
\affiliation{Diamond Light Source, Harwell Campus, Didcot, OX11 0DE, UK}

\author{J. Grant}
\affiliation{Computing Services, University of Bath, Claverton Down, Bath,
BA2 7AY, UK}

\author{A. J. Smith}
\affiliation{Diamond Light Source, Harwell Campus, Didcot, OX11 0DE, UK}

\author{J. L. Rawle}
\affiliation{Diamond Light Source, Harwell Campus, Didcot, OX11 0DE, UK}

\author{D. J. Barlow}
\affiliation{Institute of Pharmaceutical Science, King's College London,
London, SE1 9NH, UK}

\author{M. J. Lawrence}
\affiliation{Division of Pharmacy and Optometry, University of Manchester,
Manchester, M13 9PT, UK}

\author{S.~C. Parker}
\affiliation{Department of Chemistry, University of Bath, Claverton Down,
Bath, BA2 7AY, UK}

\author{K.~J. Edler}
\email{k.edler@bath.ac.uk}
\affiliation{Department of Chemistry, University of Bath, Claverton Down,
Bath, BA2 7AY, UK}

\date{\today}

\begin{abstract}
Using molecular simulation to aid in the analysis of neutron reflectometry measurements is commonplace.
However, reflectometry is a tool to probe large-scale structures, and therefore the use of all-atom simulation may be irrelevant.
This work presents the first direct comparison between the reflectometry profiles obtained from different all-atom and coarse-grained molecular dynamics simulations.
These are compared with a traditional model layer structure analysis method to determine the minimum simulation resolution required to accurately reproduce experimental data.
We find that systematic limits reduce the efficacy of the MARTINI potential model, while the Berger united-atom and Slipids all-atom potential models agree similarly well with the experimental data.
The model layer structure gives the best agreement, however, the higher resolution simulation-dependent methods produce an agreement that is comparable.
Finally, we use the atomistic simulation to advise on possible improvements that may be offered to the model layer structures, creating a more realistic monolayer model.
\begin{description}
\item[Usage]
Electronic Supplementary Information (ESI) available: All analysis/plotting scripts and figure files, allowing for a fully reproducible, and automated, analysis workflow for the work presented is available at \url{https://github.com/arm61/sim_vs_trad} (DOI: 10.5281/zenodo.2600729) under a CC BY-SA 4.0 license.
Reduced experimental datasets are available at \url{https://researchdata.bath.ac.uk/id/eprint/586}, under a CC-BY 4.0 license.
\end{description}
\end{abstract}

\maketitle                        

\section{Introduction}

Neutron and X-ray reflectometry techniques are popular in the study of layered structures, such as polyelectrolyte-surfactant mixtures \cite{llamas_study_2018}, lipid bilayer systems \cite{waldie_localization_2018}, electrodeposited films \cite{beebee_effect_2019}, and dye-sensitised solar cell materials \cite{mccree-grey_preferred_2015}.
Unlike other surface-sensitive techniques, such as atomic force microscopy (AFM) or scanning electron microscopy (SEM), reflectometry methods can investigate buried interfaces in addition to the material surface.
This is due to the ability of neutrons and X-rays to probe more deeply into a material than an AFM tip or the electron.
Additionally, reflectometry techniques can more easily provide information about the average structure over large regions of material, resulting in significantly improved sampling, compared with microscopy techniques \cite{renaud_probing_2009}.
The growth in popularity of reflectometry techniques can be attributed to the significant development of both neutron and X-ray reflectometry instrumentation, such as FIGARO, the horizontal neutron reflectometer at the ILL \cite{campbell_figaro_2011}, and the beam deflection system at the I07 beamline of the Diamond Light Source \cite{arnold_implementation_2012}.

Typically, the analysis of a neutron or X-ray reflectometry profile is achieved by the application of the Abel\`{e}s matrix formalism for stratified media \cite{abeles_sur_1948,parratt_surface_1954} to a model layer structure.
These layer structures are usually defined by the underlying chemistry of the system, for example, the chemically-consistent method that we previously used \cite{mccluskey_bayesian_2019}, which accounts for the chemical linkage between the phospholipid head and tail layers.
However, there has been growing interest in the use of molecular dynamics simulations to inform the development of these layer structures.
This is due to the fact that the equilibrium structures for soft matter interfaces, that are often of interest in reflectometry studies, are accessible on all-atom simulation timescales \cite{scoppola_combining_2018}.
However, to the authors' knowledge, no work has directly compared different levels of simulation coarse-graining in order to assess the required resolution for the accurate reproduction of a given neutron reflectometry profile.

The use of MD-driven analysis of neutron reflectometry usually involves, either the calculation of the SLD profile from the simulation or the full determination of the reflectometry profile.
In the former case, the calculated SLD profile may be compared with the SLD profile determined from the use of a model layer structure analysis method.
Bobone \emph{et al.} used such a method to study the antimicrobial peptide trichogin GA-IV within a supported lipid bilayer \cite{bobone_membrane_2013}.
A four layer-model consisted of the hydrated \ce{SiO2} layer, an inner lipid head-region, a lipid tail-region, and an outer lipid head region.
The SLD profile from the MD simulations agreed well with that fitted to the reflectometry data from this model layer structure.

The reflectometry profile was calculated explicitly from classical simulation in the works of Miller \emph{et al.} and Anderson and Wilson \cite{miller_monte_2003,anderson_molecular_2004}.
In these, an amphiphilic polymer at the oil-water interface was simulated by Monte Carlo and MD respectively, and the neutron reflectometry profile found by splitting the simulation cell into a series of small layers and applying the Abel\`{e}s matrix formalism.
There was good agreement between the experimental and calculated reflectometry, for low interface coverages of the polymer.
Another study that has made a direct comparison between the atomistic simulation-derived reflectometry and those measured experimentally is that of Darr\'{e} \emph{et al.} \cite{darre_molecular_2015}.
Darr\'{e} \emph{et al.}, NeutronRefTools was developed to produce the neutron reflectometry profile directly from an MD simulation.
The particular system studied was a supported 1,2-dimyristoyl-\emph{sn}-glycero-3-phosphocholine (DMPC) lipid bilayer, again good agreement was found between the simulation-derived profile and the experimental measurement.
However, the nature of the support required a correction for the head-group hydration to be imposed to achieve this agreement.

Koutsioubas used the MARTINI coarse-grained representation of a 1,2-dipalmitoyl-\emph{sn}-glycero-3-phosphocholine (DPPC) lipid bilayer to compare with experimental reflectometry \cite{koutsioubas_combined_2016}.
This work showed that the parameterisation of the MARTINI water beads was extremely important in the reproduction of the reflectometry data, as the non-polarisable water bead would freeze into crystalline sheets resulting in artefacts in the reflectometry profiles calculated.
The work of Hughes \emph{et al.} studied again a DPPC lipid bilayer system \cite{hughes_interpretation_2016}, albeit an all-atom representation, that was compared with a supported DPPC lipid bilayer system measured with polarised neutron reflectometry.
The SLD profile found from MD was varied to better fit the experimental measurement, resulting in good agreement.
Additionally, the ability to vary the SLD profile was used to remove artefacts that arose when the MD simulations were merged with the Abel\`{e}s matrix formalism.
This was done to account for regions present in the experiment that were not modelled explicitly.

In all of the examples discussed so far there is no direct comparison between the reflectometry profile determined from simulation and that from the application of a traditional analysis method.
Indeed, the only example, to the authors' knowledge where a direct comparison was drawn is the work of Dabkowska \emph{et al.} \cite{dabkowska_modulation_2014}.
This work compares the reflectometry profile from a DPPC monolayer at the air-water interface containing dimethyl sulfoxide molecules with a similar molecular dynamics simulation parameterised with the CHARMM potential model.
The use of multimodal analysis allowed the determination of the position of a concentration of DMSO molecules at a particular region within a monolayer and the orientation of such molecules.

The previously mentioned work of Koutsioubas involved the use of the MARTINI coarse-grained force field to simulations the DPPC bilayer system \cite{koutsioubas_combined_2016}.
The use of atomistic simulation for soft matter systems, such as a lipid bilayer, is undesirable as this requires a huge number of atoms to be simulated, due to the large lengths scales involved.
The purpose of simulation coarse-graining is to reduce the number of particles over which the forces must be integrated, additionally by removing the higher frequency bond vibrations, the simulation timestep can also be increased \cite{pluhackova_biomembranes_2015}.
Together, these two factors enable an increase in both simulation size and length.
The use of the MARTINI 4-to-1 coarse-grained and the Berger united-atom (where hydrogen atoms are integrated into the heavier atoms to which they are bound) potential models are particularly pertinent for application to lipid simulations as both were developed with this specific application in mind \cite{marrink_martini_2007,berger_molecular_1997}.

The MARTINI potential model involves integrating the interactions of every four heavy atoms, i.e. those larger than hydrogen, into beads of different chemical nature.
This potential model attempts to simplify the interactions of lipid and protein molecules significantly by allowing for only eighteen particle types, defined by their polarity, charge, and hydrogen-bond acceptor/donor character, which are discussed in detail in the work of Marrink \emph{et al.} \cite{marrink_martini_2007}.
Increasing the simulation resolution gives an united-atom potential mode, where all of the hydrogen atoms are integrated into the heavier atoms to which they are bound.
One of the most popular united-atom potential models for lipid simulations is that developed by Berger \emph{et al.} \cite{berger_molecular_1997}, with the original paper being cited 1500 times at the time of writing.
Finally, the all-atom Slipid (Stockholm Lipids) lipid potential model was developed in 2012 by J\"{a}mbeck and Lyubartsev \cite{jambeck_derivation_2012}.
All three of these potential models were designed to model lipid bilayer systems.

It is clear that there is substantial interest in the use of classical simulation, and coarse-graining for the analysis of neutron reflectometry data.
However, there has been no work to investigate whether the use of atomistic simulations gives more detailed than is required to reproduce the reflectometry profile accurately or to assess whether the application of a coarse-grained representation is suitable to aid in analysis.
In this work, three potential models, with different degrees of coarse-graining; namely the Slipid all-atom \cite{jambeck_derivation_2012}, Berger united-atom \cite{berger_molecular_1997}, and MARTINI coarse-grained potential models \cite{marrink_martini_2007}, are compared in terms of their ability to reproduce neutron reflectometry data.
We consider that this work offers a fundamental insight into the potential model resolution that is necessary to accurately reproduce experimental neutron reflectometry measurements.
Furthermore, we use the highest resolution simulations to suggest possible adjustments that may be made to the model layer structure analysis methods that are typically used for the rationalisation of neutron reflectometry.

\section{Methodology}

\subsection{Neutron reflectometry measurements}
The neutron reflectometry measurements analysed in this work have been previously published by Hollinshead \emph{et al.} \cite{hollinshead_effects_2009} and full details of the experimental methods can be found in this previous publication.
These measurements concern the study of a monolayer of 1,2-distearoyl-\emph{sn}-phosphatidylcholine (DSPC) at the air-water interface.
The neutron reflectometry measurements were conducted on seven isotopic contrasts of the lipid and water.
These contrasts were made up from four lipid types; fully-hydrogenated lipid (h-DSPC), head-deuterated lipid (\ce{d_{13}}-DSPC), tail-deuterated lipid (\ce{d_{70}}-DSPC), and fully-deutered lipid (\ce{d_{83}}-DSPC), were paired with two water contrasts; fully-deuterated water \ce{D2O} and air-contrast matched water (ACMW), where \ce{D2O} and \ce{H2O} are
mixed such that the SLD is zero.
The pairing of the fully-hydrogenated lipid with ACMW was not used due to the lack of scattering available from such a system.
Measurements were conducted at four different surface pressures; \SIlist{20;30;40;50}{\milli\newton\per\meter}.
Table~\ref{tab:nom} outlines the shorthands used to refer to the different contrast pairings in this work.
\begin{table}[h]
\small
  \caption{\ The different contrasts of lipid and water investigated in this
  work.}
  \label{tab:nom}
  \begin{tabular*}{0.48\textwidth}{@{\extracolsep{\fill}}lll}
    \hline
    Shorthand & Lipid contrast & Water contrast \\
    \hline
    h-\ce{D2O} & h-DSPC & \ce{D2O} \\
    \ce{d_{13}}-ACMW & \ce{d_{13}}-DSPC & ACMW \\
    \ce{d_{13}}-\ce{D2O} & \ce{d_{13}}-DSPC & \ce{D2O} \\
    \ce{d_{70}}-ACMW & \ce{d_{70}}-DSPC & ACMW \\
    \ce{d_{70}}-\ce{D2O} & \ce{d_{70}}-DSPC & \ce{D2O} \\
    \ce{d_{83}}-ACMW & \ce{d_{83}}-DSPC & ACMW \\
    \ce{d_{83}}-\ce{D2O} & \ce{d_{83}}-DSPC & \ce{D2O} \\
    \hline
  \end{tabular*}
\end{table}

\subsection{Molecular dynamics simulations}
The DSPC monolayer simulations were made up of lipid molecules modelled with three potential models, each of a different particle grain-size.
The Slipids potential model is an all-atom representation of the lipid molecules \cite{jambeck_derivation_2012}, which was used alongside the single point charge (SPC) water model \cite{berendsen_missing_1987}, with a timestep of \SI{0.5}{\femto\second}, the SHAKE, RATTLE, and PLINCS methods were used to constrain the \ce{C-H} bonds \cite{miyamoto_settle_1992,hess_p-lincs_2008}.
The Berger potential model is obtained by the integration of the hydrogen atoms into the heavy atoms to which they are bound, producing a united-atom potential model \cite{berger_molecular_1997}; again the SPC water model was used.
This potential model was simulated with an increased timestep of \SI{1}{\femto\second}.
It is noted that these timesteps are shorter than those typically used for both forcefields, and timesteps of up to \SI{2}{\femto\second} have been applied previously \cite{berger_molecular_1997,jambeck_derivation_2012}.
Finally, the lowest resolution potential model used was the MARTINI \cite{marrink_martini_2007} alongside the polarisable MARTINI water model \cite{yesylevskyy_polarizable_2010}, to avoid the freezing issues observed previously \cite{koutsioubas_combined_2016}.
The MARTINI 4-to-1 heavy atom beading allows for the use of a \SI{20}{\femto\second} timestep.
For the Slipids and Berger potential model a short-range cut-off of \SI{10}{\angstrom} was used, while for the MARTINI potential model the cut-off was extended to \SI{15}{\angstrom}.
All simulations were conducted with temperature coupling to a heat bath at \SI{300}{\kelvin} and a leap-frog integrator, and run using GROMACS 5.0.5 \cite{berendsen_gromacs_1995,lindahl_gromacs_2001,van_der_spoel_gromacs_2005,hess_gromacs_2008} on 32 cores of the STFC Scientific Computing resource SCARF.
The simulation was of a monolayer, therefore the Ewald 3DC correction was applied to allow for the use of \emph{x}/\emph{y}-only periodic boundary conditions \cite{yeh_ewald_1999}.
A close-packed ``wall'' of non-interacting dummy atoms was placed at each side of the simulation cell in the \emph{z}-direction to ensure that the atoms could not leave the simulation cell.

\begin{figure}[h]
\centering
  \includegraphics[width=0.4\textwidth]{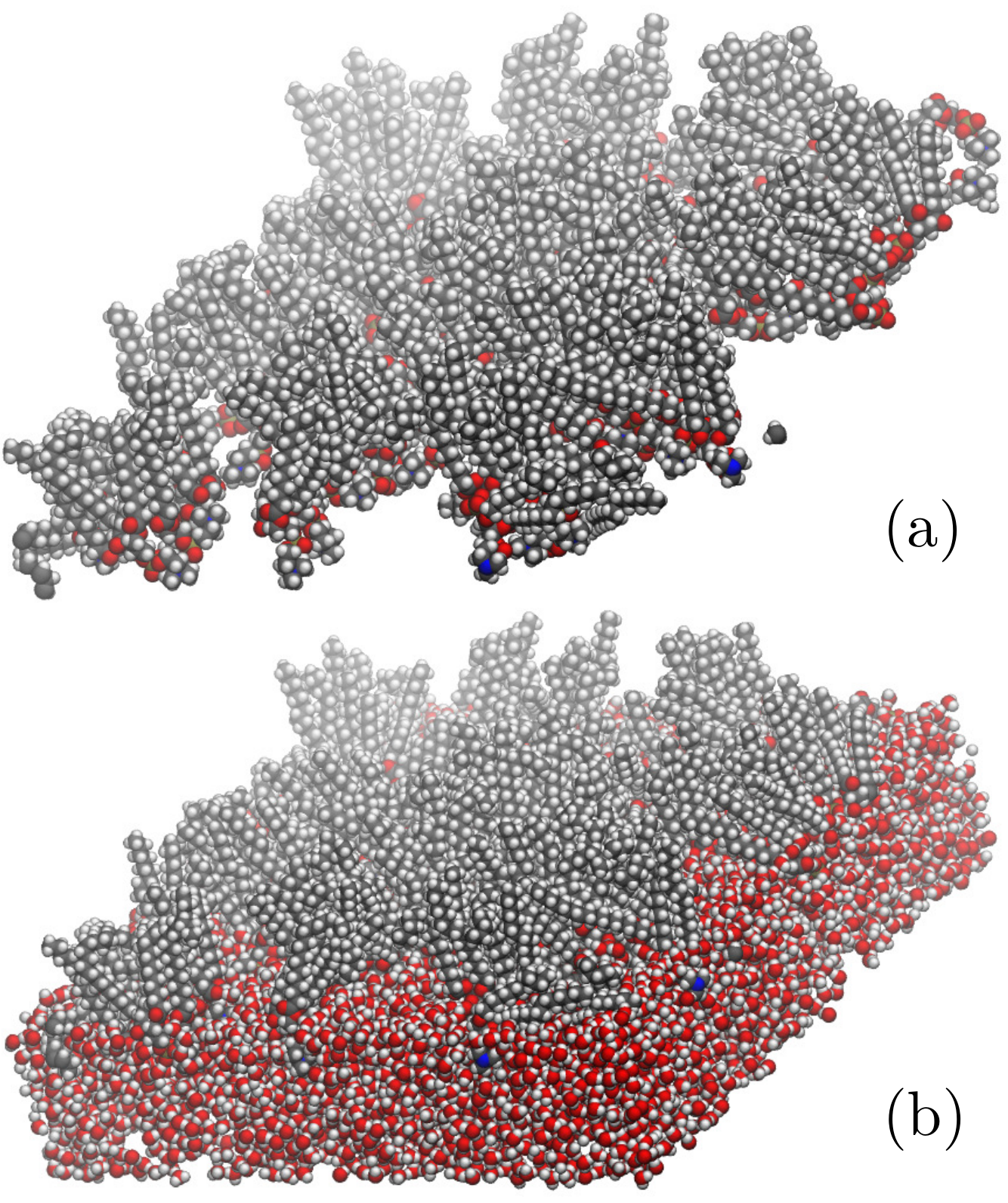}
  \caption{The DSPC monolayer (a) without water layer and (b) with water
  layer, visuallised using VMD\cite{humphrey_vmd_1996}.}
  \label{fig:drywet}
\end{figure}

\begin{figure}[h]
\centering
  \includegraphics[width=0.48\textwidth]{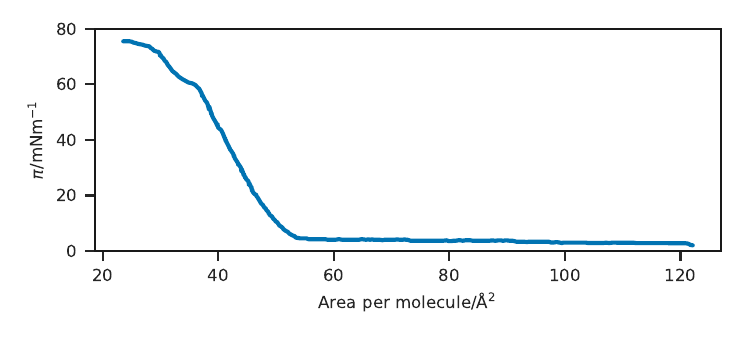}
  \caption{The experimental surface pressure isotherm for DSPC on water, taken from
  the work of Kubo \emph{et al.}\cite{kubo_phosphatidylcholine_2001}.}
  \label{fig:iso}
\end{figure}

The starting simulation structure was generated using the molecular packing software Packmol \cite{martinez_packmol_2009}.
This was used to produce a monolayer of 100 DSPC molecules, with the head groups oriented to the bottom of the simulation cell.
A \SI{6}{\angstrom} layer of water was then added such that it overlapped the head groups, this was achieved using the \texttt{solvate} functionality in GROMACS 5.0.5.
Examples of a dry and a wet monolayer can be seen in Figure~\ref{fig:drywet} for the Berger potential model representation.
A general protocol was then used to relax the system at the desired surface coverage, reproducing the effects of a Langmuir trough \emph{in silico}.
This involved subjecting the system to a semi-isotropic barostat, with a compressibility of \SI{4.5E-5}{\per\bar} for the Slipids and Berger simulations and \SI{3.0E-4}{\per\bar} for the MARTINI simulations.
The pressure in the \emph{z}-dimension was kept constant at \SI{1}{\bar}, while it was increased in the \emph{x}- and \emph{y}-dimensions isotropically.
This allowed for the surface area of the interface to reduce, as the lipid molecules have a preference to stay at the interface, while the total volume of the system stayed relatively constant, as the water molecules moved down to relax the pressure in the \emph{z}-dimension.
When the \emph{xy}-surface area is reached that is associated with the area per molecule (APM) for each surface pressure, described by the experimental surface pressure-isotherm (Figure~\ref{fig:iso}), given in Table~\ref{tbl:apm}, the coordinates were saved and used as the starting structure for the equilibration simulation.
This equilibration simulation involved continuing the use of the semi-isotropic barostat, with the \emph{xy}-area of the box fixed, allowing the system to relax at a pressure of \SI{1}{\bar} in the \emph{z}-dimension.
Following the application of the pair of semi-isotropic barostats, the thickness of the water layer was typically in the region of \SI{30}{\angstrom}.
The equilibration period was \SI{1}{\nano\second}, following which the \SI{50}{\nano\second} NVT ensemble production simulations were run, on which all analyses were conducted.
\begin{table}[h]
\small
  \caption{\ The areas per molecule (APM) associated with particular surface
  pressures and the size of the \emph{x}- and \emph{y}-cell dimension for a
  simulation of 100 lipid molecules.}
  \label{tbl:apm}
  \begin{tabular*}{0.48\textwidth}{@{\extracolsep{\fill}}lll}
    \hline
    $\pi$/\si{\milli\newton\per\meter} & APM/\si{\angstrom\squared} &
    \emph{xy}-cell length/\si{\angstrom} \\
    \hline
    20 & 47.9 & 69.1 \\
    30 & 46.4 & 68.1 \\
    40 & 45.0 & 67.1 \\
    50 & 44.6 & 66.0 \\
    \hline
  \end{tabular*}
\end{table}

\subsection{Abel\`{e}s matrix formalism}
To compare with the simulation-derived reflectometry profiles, a modified version of the chemically-consistent surfactant monolayer model previously used in the group was applied \cite{mccluskey_bayesian_2019,mccluskey_lipids_at_airdes_2019}.
This model is implemented as a class that is compatible with the Python package \texttt{refnx} \cite{nelson_refnx_2019,nelson_refnx_2018} and is made up of two layers; the head-layer at the interface with the solvent and the tail-layer at the interface with the air.
The head components have a calculated scattering length, $b_h$, (found as a summation of the neutron scattering lengths of the individual atoms, see Table S1 of the ESI) and a component volume, $V_h$.
These make up a head-layer with a given thickness, $d_h$, and interfacial roughness, $\sigma_h$, and within this layer, some volume fraction of solvent may intercalate, $\phi_h$.
The tail components also have a similarly calculated scattering length, $b_t$, and component volume, $V_t$.
This tail-layer also has a given thickness, $d_t$, and interfacial roughness, $\sigma_t$.
A maximum value for the thickness of the tail-layer was imposed, this value was taken from the Tanford Equation \cite{tanford_hydrophobic_1980},
\begin{equation}
  t_t = 1.54 + 1.265n,
\end{equation}
where $n$ is the number of carbon atoms in the chain, and so for DSPC $t_t = \SI{24.3}{\angstrom}$.
The SLD of the tail and head layers used in the Abel\`{e}s matrix formalism can, therefore, be found as,
\begin{equation}
  \text{SLD}_i = \frac{b_i}{V_i}(1 - \phi_i) + \text{SLD}_s(\phi_i),
\end{equation}
where, $\text{SLD}_s$ is the scattering length density of the subphase (water), and $i$ indicates either the tail- or head-layer; it is assumed that the tail layer contains no solvent or air, i.e. $\phi_t = 0$ in agreement with the work of Campbell \emph{et al.} \cite{campbell_structure_2018}.
To ensure that the number density of the head components and pairs of tail components is the same, the following constraint was included in the model \cite{braun_polymers_2017},
\begin{equation}
  \phi_h = 1 - \bigg(\frac{d_tV_h}{V_td_h}\bigg).
\end{equation}
A single value for the interfacial roughness was fitted for all interfaces, which was limited to be no less than \SI{3}{\angstrom}, as there is only a single lipid type in each monolayer \cite{campbell_structure_2018}.
Therefore, any roughness at the air-water interface is carried equally through all the layers, in a conformal fashion \cite{kozhevnikov_general_2012}.
The modifications over the previous implementation were that the tail component volume was constrained, based on the APM (taken from the surface pressure isotherm),
\begin{equation}
  V_t = d_t \text{APM},
\end{equation}
resulting in the monolayer model and simulation-derived models being equally constrained by the calculated surface coverage.
Additionally, the head component volume was constrained to a value of \SI{339.5}{\angstrom\cubed}, in agreement with the work of Ku\v{c}erka \emph{et al.}\cite{kucerka_determination_2004} and Balgav\'{y} \emph{et al.} \cite{balgavy_evaluation_2001}.
A uniform background, limited to lie within \SI{10}{\percent} of the highest $q$-value reflected intensity, and a scale factor were then determined using \texttt{refnx} to offer the best agreement between the calculated reflectometry profile and that measured experimentally.

In this work, the experimental data from all seven contrasts were co-refined to a single monolayer model, where the head thickness, tail thickness, and interfacial roughness were allowed to vary.
The values of the head and tail scattering lengths, along with the super and subphase SLDs are given in Table S1.
For each co-refinement of seven neutron reflectometry measurements, there were in total five degrees of freedom in the fitting process, and the fitting was performed using a differential evolution algorithm, which has been shown to be particularly useful in the analysis of reflectometry data \cite{wormington_characterization_1999,bjorck_fitting_2011}.
To obtain uncertainties on the fitted model, Markov chain Monte Carlo sampling, enabled by the emcee package \cite{foreman-mackey_emcee_2013} was used to assess the probability distribution function for each parameter.
In the MCMC sampling, 200 walkers were used over 1000 iterations, following equilibration of 200 iterations.
The use of MCMC sampling allowed for Bayesian inference of the PDF for each of the variables and their respective interactions and the Shapiro test to be used to assess if each PDF was normally distributed.
Parameters that were shown to be normally distributed are given with symmetric confidence intervals, while those that failed the Shapiro test are given with asymmetric confidence intervals (\SI{95}{\percent} confidence intervals in both cases).
The Abel\`{e}s matrix formalism was used to calculate the reflectometry profiles as described in the ESI.

\subsection{Simulation-derived analysis}
The ESI also includes a Python class that is compatible with \texttt{refnx} \cite{nelson_refnx_2019,nelson_refnx_2018} allowing for simulation-derived reflectometry profiles to be obtained, using a similar method to that employed in previous work, such as Dabkowska \emph{et al.} \cite{dabkowska_modulation_2014}.
The Abel\`{e}s matrix formalism is applied to layers, the SLD of which is drawn directly from the simulation, and the thickness of which is defined.
The layer thickness used was \SI{1}{\angstrom} for the Slipid and Berger potential model simulations, with an interfacial roughness between these layers is defined as \SI{0}{\angstrom}.
For the MARTINI potential model, a layer thickness of \SI{4}{\angstrom} was used, with an interfacial roughness of \SI{0.4}{\angstrom}, a detailed discussion for the rationale behind this is available in the ESI.
Each of the \SI{50}{\nano\second} production simulations were analysed with a frequency of \SI{10}{\per\nano\second}, and the SLD profiles were determined by summing the scattering lengths, $b_j$, for each of the atoms in a given layer.
\begin{equation}
  \text{SLD}_n = \frac{\sum_j{b_j}}{V_n},
\end{equation}
where, $V_n$ is the volume of the layer $n$, obtained from the simulation cell parameters in the plane of the interface and the defined layer thickness.
Again a uniform background, limited to lie within \SI{10}{\percent} of the highest $q$-value reflected intensity, and a scale factor were then determined using \texttt{refnx}.

\subsection{Comparison between monolayer model and simulation-derived analysis}
\label{sec:para}
The agreement between the models from each method was assessed using the following goodness-of-fit metric, following the transformation of the data into $Rq^4$ space,
\begin{equation}
  \chi^2 = \sum_{i=1}^{N_{\text{data}}} \frac{[R_{\text{exp}}(q_i) -
  R_{\text{sim}}(q_i)]^2}{[\delta R_{\text{exp}}(q_i)]^2},
\end{equation}
where $q_i$ is a given $q$-vector, which depends on the neutron wavelength and reflected angle, $R_{\text{exp}}(q_i)$ is the experimental reflected intensity, $R_{\text{sim}}(q_i)$ is the simulation-derived reflected intensity, and $\delta R_{\text{exp}}(q_i)$ is the resolution function of the data.

The number of water molecules per head group, wph, was also compared between the different methods.
This was obtained from the chemically-consistent model by considering the solvent fraction in the head-layer, $\phi_h$, the volume of the head group, $V_h$, and taking the volume of a single water molecule to be \SI{29.9}{\cubic\angstrom} (from the density of water as
\SI{997}{\kilo\gram\per\cubic\meter}),
\begin{equation}
  \text{wph} = \frac{\phi_hV_h}{29.9 - 29.9\phi_h}.
  \label{equ:wph}
\end{equation}
The number densities, in the \emph{z}-dimension, for each of the three components (lipids heads, tails, and water) may be obtained directly from the MD simulation trajectory.
In order to determine the number of water molecules per headgroup from the MD simulations, a head-layer region was defined as that which contained \SI{60}{\percent} of the lipid head number density.
The ratio between the water density and the lipid head density was then found within this head-layer region.

\subsection{Simulation trajectory analysis}
\label{sec:traj}
In order to use the MD trajectory to guide the future development of the chemically-consistent layer model, it was necessary to investigate the solvent penetration into the head group region of the lipids, the roughness of each interface and the lipid tail length.
The solvent penetration was determined using the intrinsic surface approach, as detailed by Allen \emph{et al.} \cite{allen_specific_2016,pandit_algorithm_2003}.
The intrinsic surface approach enables the calculation of the solvent penetration without the effect of the monolayer roughness.
This involves taking the \emph{z}-dimension position of each water molecule with respect to an anchor point, in this work the anchor point was the phosphorus atom of the lipid head that was closest to the water molecule in the \emph{xy}-plane.
The roughness was probed by investigating the variation in positions for the start, middle, and end of each of the head and tail groups.
The start of the lipid head was defined as the nitrogen atom, the middle the phosphorus and the end the tertiary carbon, while the start of the lipid tail was defined as the carbonyl carbon atom, the middle the ninth carbon in the tail and the end the final carbon atom in the tail.
The distribution of each of these atom types was determined by finding the \SI{95}{\percent} quantile for the position in the \emph{z}-dimension and comparing the spread of the mean and the upper quantile.
Finally, the tail length distance, $t_t$ was found as the distance from the carbonyl carbon atom to the final primary carbon atom of the lipid tail.
All of these analyses used \texttt{MDAnalysis} package \cite{gowers_mdanalysis_2016,michaud-agrawal_mdanalysis_2011} and the scripts that were used can be found in the ESI.

\section{Results \& Discussion}
\begin{figure*}
 \centering
 \includegraphics[width=0.49\textwidth]{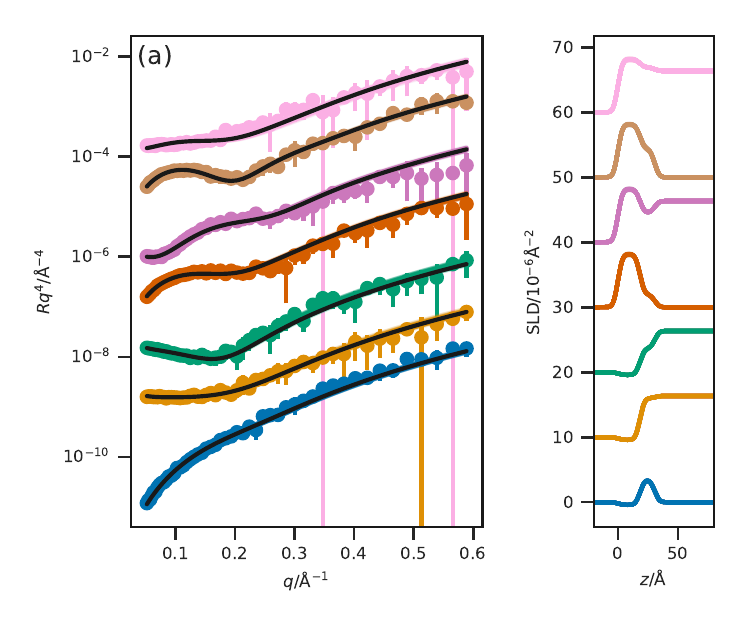}
 \includegraphics[width=0.49\textwidth]{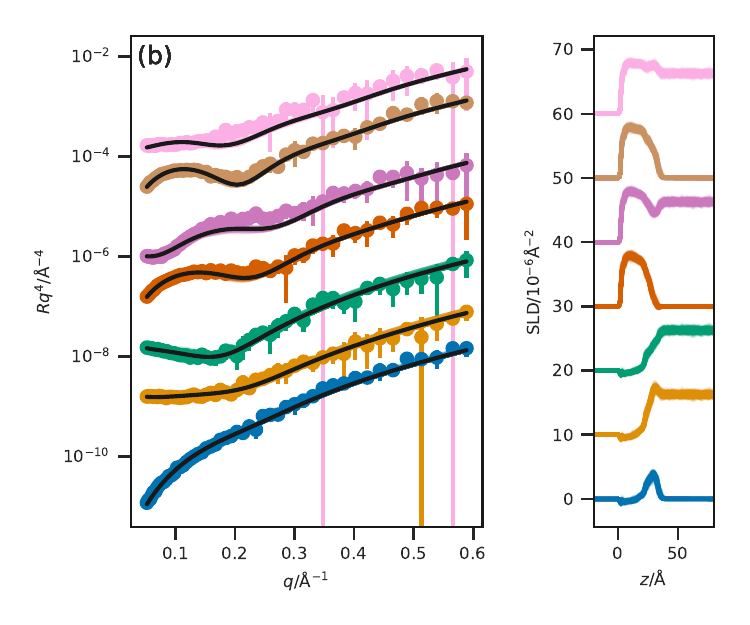} \\
 \includegraphics[width=0.49\textwidth]{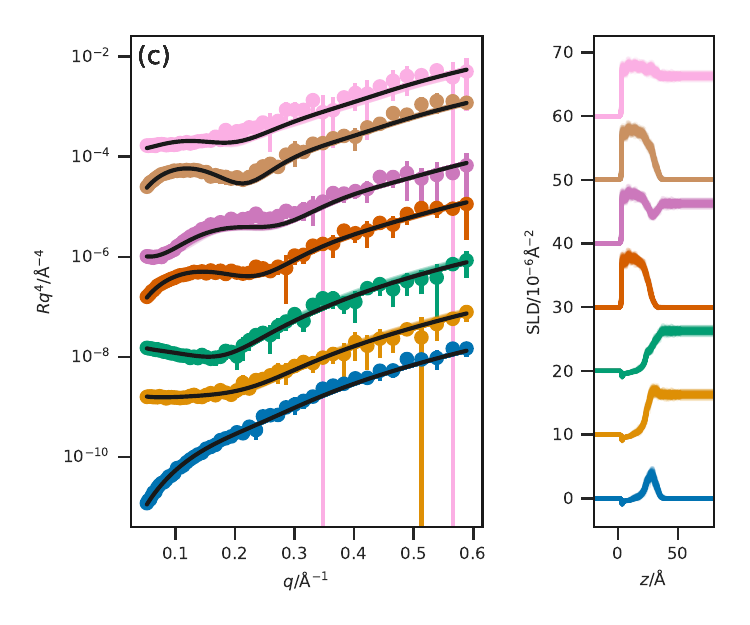}
 \includegraphics[width=0.49\textwidth]{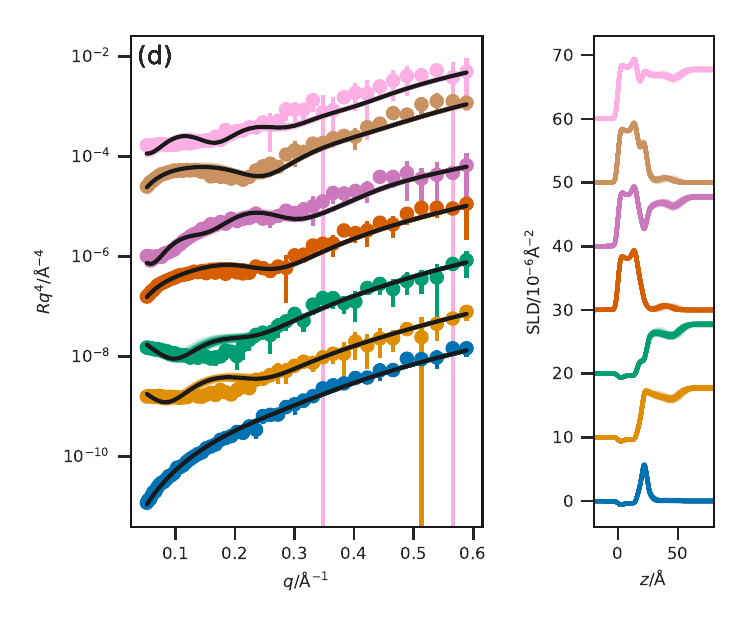}
 \caption{A comparison of the reflectometry and SLD profiles obtained from
 (a) the chemically-consistent layer model, (b) the Slipid simulation, (c) the Berger
 simulation, and (d) the MARTINI simulation, at an APM associated with a surface pressure of \SI{30}{\milli\newton\per\meter}. From top-to-bottom the contrasts are as
 follows; \ce{d_{83}}-\ce{D2O}, \ce{d_{83}}-ACMW, \ce{d_{70}}-\ce{D2O},
 \ce{d_{70}}-ACMW, h-\ce{D2O}, \ce{d_{13}}-\ce{D2O}, \ce{d_{13}}-ACMW.
 The different contrast reflectometry profiles have been offset in the
 \emph{y}-axis by an order of magnitude and the SLD profiles offset in
 the \emph{y}-axis by \SI{10e-6}{\per\square\angstrom}, for clarity.}
 \label{fig:ref}
\end{figure*}
Figure~\ref{fig:ref} presents the reflectometry and SLD profiles from each of the different methods, both the traditional layer model and the three potential model simulations, at an APM associated with a surface pressure of \SI{30}{\milli\newton\per\meter}.
This work will focus discussions on the data at this surface pressure, however other surface pressures showed similar trends and can be found in the ESI.
In addition, the $\chi^2$ for each contrast, average $\chi^2$, and standard deviation for each method are given in Table~\ref{tbl:chi} for each contrast.
\begin{table*}
\small
  \caption{\ The goodness-of-fit between the calculated and experimental
  reflectometry profile at a surface pressure of
  \SI{30}{\milli\newton\per\metre}.}
  \label{tbl:chi}
  \begin{tabular*}{\textwidth}{@{\extracolsep{\fill}}lllll}
    \hline
    Contrast & Monolayer model & Slipid & Berger & MARTINI \\
    \hline
    h-\ce{D2O} & 37.82 &
    154.91 &
    107.69 &
    1427.77 \\
    \ce{d_{13}}-ACMW & 74.39 &
    79.71 &
    72.39 &
    124.33 \\
    \ce{d_{13}}-\ce{D2O} & 36.10 &
    225.26 &
    87.18 &
    1987.96 \\
    \ce{d_{70}}-ACMW & 112.91 &
    74.30 &
    91.28 &
    345.40 \\
    \ce{d_{70}}-\ce{D2O} & 183.04 &
    622.06 &
    549.92 &
    1873.57 \\
    \ce{d_{83}}-ACMW & 78.36 &
    134.18 &
    315.34 &
    706.97 \\
    \ce{d_{83}}-\ce{D2O} & 273.28 &
    331.01 &
    418.31 &
    3128.58 \\
    \hline
    Average$\pm$Standard deviation &
    $113.70\pm80.04$ &
    $231.63\pm240.29$ &
    $234.59\pm232.66$ &
    $1370.66\pm1046.56$ \\
    \hline
  \end{tabular*}
\end{table*}
\subsection{Traditional analysis}
The chemically-consistent model was used to determine the structure of the lipid monolayer, Table~\ref{tab:cc} gives the optimum values for the parameters that were varied in the model.
It is clear from this Table, that as the surface pressure is increased, as expected (and as found previously \cite{mohwald_phospholipid_1990,vaknin_structural_1991}), the overall thickness of the monolayer increases.
The thickness increase for the lipid tails may be associated with the straightening of the tails with respect to the interface normal, while the thickness increase of the head groups has been noted previously for DSPC \cite{hollinshead_effects_2009}.
\begin{table*}
\small
  \caption{\ The values for the parameters allowed to vary in the fitting of the chemically-consistent model, at each surface pressure measured.}
  \label{tab:cc}
  \begin{tabular*}{\textwidth}{@{\extracolsep{\fill}}llllll}
    \hline
    Surface Pressure/\si{\milli\newton\per\meter} & $d_h$/\si{\angstrom} & $d_t$/\si{\angstrom} & $\sigma_{t,h,s}$/\si{\angstrom} & $\phi_h$$\times10^{-2}$ & $V_t$/\si{\angstrom\cubed} \\
    \hline
    20 & $11.00^{+0.48}_{-0.49}$ & $18.20\pm{0.23}$ & $3.02^{+0.08}_{-0.02}$ & $35.57\pm{2.99}$ & $871.67^{+11.27}_{-11.02}$ \\
    30 & $12.27\pm{0.49}$ & $18.33\pm{0.24}$ & $3.01^{+0.07}_{-0.01}$ & $40.38^{+2.29}_{-2.47}$ & $850.38\pm{10.95}$ \\
    40 & $13.54\pm{0.49}$ & $18.60^{+0.22}_{-0.22}$ & $3.03^{+0.12}_{-0.03}$ & $44.30\pm{2.10}$ & $836.95\pm{9.88}$ \\
    50 & $14.27\pm{0.46}$ & $19.20^{+0.22}_{-0.27}$ & $3.10^{+0.22}_{-0.10}$ & $46.68\pm{1.77}$ & $856.25\pm{12.08}$ \\
    \hline
  \end{tabular*}
\end{table*}

It would be anticipated that as the surface pressure increases, there would be a corresponding decrease in the volume fraction of solvent in the head group \cite{bayerl_specular_1990}.
However, for DSPC, the volume fraction of the solvent appears to be constant (or even increase slightly) with increasing surface pressure.
We believe that this is due to the decision to constrain the volume of the lipid head, which may decrease with increasing surface pressure.
It has been noted previously that the interfacial roughness will increase with increasing surface pressure \cite{lu_aspects_1994}, this can be observed with the slight increase between \SIrange{20}{50}{\milli\newton\per\meter}.

Hollinshead \emph{et al.} \cite{hollinshead_effects_2009} suggest a tail volume of \SI{972}{\angstrom\cubed} from the density data.
However, the values found in this work are substantially lower, at \SI{\sim850}{\angstrom\cubed}.
This reduction, of \SI{\sim12}{\percent}, agrees well with the work of Campbell \emph{et al.} \cite{campbell_structure_2018} and Small \cite{small_lateral_1984}, which suggest that under the surface pressure investigated in this work a reduction of the tail volume of up to \SI{15}{\percent} may be observed.
We believe that the model layer structure from the chemically-consistent method provides a satisfactory description of the monolayer structure.
However, the use of an MD-driven analysis method may provide greater insight into the chemical nature of the monolayer.

\subsection{MARTINI}
It is clear from Figure~\ref{fig:ref} and Table~\ref{tbl:chi}, that the MARTINI potential model simulations do not effectively reproduce the reflectometry profile, with a clear difference between the model and data.
The SLD profiles derived from the MARTINI simulations contain significant dislocations, which lead to artefacts in the resulting reflectometry profile, and therefore the poor agreement with the data.

It is noted that the agreement with the contrasts containing \ce{D2O} is particularly poor.
This is most likely an artefact of the structuring effect from the wall at the bottom of the simulation cell on the polarisable MARTINI water.
It is noted that this may be reduced through the use of a less-ordered wall structure \cite{koutsioubas_combined_2016}.
Alternatively, it may be possible to completely remove the presence of this structuring through the inclusion of \SI{\sim10}{\percent} of antifreeze MARTINI beads alongside the normal MARTINI water.
However, this method has been noted to also give structuring effects in the presence of ordered walls \cite{marrink_comment_2010}.

Another artefact present in the MARTINI potential model simulations, particularly notable in the \ce{d_{83}}-ACMW and \ce{d_{70}}-ACMW contrasts where the reflectometry fringe at low-$q$ is substantially broader than represented in the data, is that the length of the hydrocarbon tail in the simulation was found to be $16.60^{+1.65}_{-1.88}$~\si{\angstrom}.
This is significantly less than the \SI{24.3}{\angstrom} estimated by the Tanford equation.
The reduction in the tail length is due to the nature of the MARTINI's 4-to-1 beading process, as DSPC has a hydrocarbon tail consisting of 18 carbon atoms, and it is not possible to bead such a chain accurately with the MARTINI potential model.
In this work, a MARTINI lipid molecule was used with 4 MARTINI beads making up the chain; corresponding to an all-atom hydrocarbon chain of 16 atoms.
Applying the Tanford equation to a hydrocarbon chain of such a length results in an anticipated length of \SI{18.7}{\angstrom}, which agrees better with that found from the simulation.

The requirement for a 4-to-1 beading structure of the MARTINI potential model is a significant weakness in the utility of this potential model in this work.
A better method may be limiting experiments to systems that can be modelled exactly or the use of a 2-to-1 beading model.
However, we are not aware of an off-the-shelf 2-to-1 coarse-grained potential model that is commonly applied to lipid molecules.

\subsection{Comparison of other simulations}
Table~\ref{tbl:chi} shows that both the Slipid and Berger potential models agree well with the experimental data, with small values for the $\chi^2$.
While Figure~\ref{fig:ref} shows that the SLD profiles both appear qualitatively similar to those from the model layer structure method.
Furthermore, the quality of agreement between these higher-resolution potential models and the model layer structure is relatively similar.
However, the model layer structure still offers a better fit to the experimental data than those determined from MD simulation.

The result that the model layer structure offers better agreement with the data than those from even all-atom simulation is to be expected, simply by considering the level of constraint present implicitly when determining the reflectometry profile directly from a simulation.
While the model layer structure constrains the layer model to be chemically-consistent, those from MD simulation have real chemical constraints present in the simulation; e.g. the bonding of atoms, and the non-bonded potentials.
The quality of the agreement from this multi-modal analysis technique is sufficient for such a method to be applied regularly in the analysis of neutron reflectometry.

Both the Slipid and Berger simulations produced values for the tail length that were in better agreement with the Tanford equation than the MARTINI simulation.
For the Slipid simulation, the tail length was found to be $20.17^{+1.41}_{-7.39}$~\si{\angstrom}, while for the Berger simulations a value of $19.80^{+1.59}_{-8.17}$~\si{\angstrom} was obtained.
Neither is quite as large as the \SI{24.3}{\angstrom} from the Tanford equation, however, it should be noted that this value is considered a maximum for the \emph{fully extended} carbon tail.

Using the molecular dynamics simulations, and the model layer structure it is possible to compare the number of water molecules per head group.
From the Slipids and Berger simulations, the number of water molecules per head group was found to be $6.41^{+1.63}_{-0.76}$ and $5.49^{+0.68}_{-0.53}$ respectively.
These are in good agreement with the $7.69\pm{0.76}$ found from the monolayer model method in conjunction with Equation \ref{equ:wph}.

\begin{figure}
 \centering
 \includegraphics[width=0.49\textwidth]{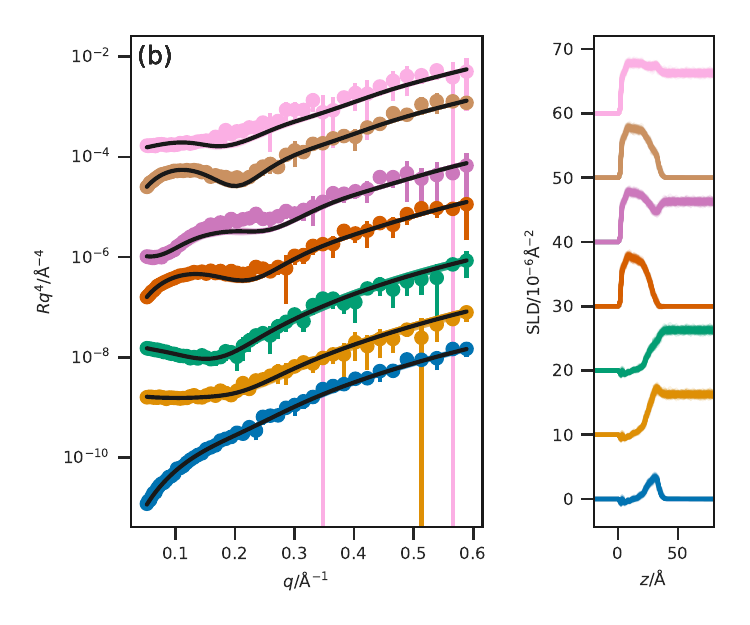}
 \caption{The reflectometry and SLD profiles obtained from the first \SI{5}{\nano\second} of the Slipid potential model simulation, at an APM associated with a surface pressure of \SI{30}{\milli\newton\per\meter}. From top-to-bottom the contrasts are as follows; \ce{d_{83}}-\ce{D2O}, \ce{d_{83}}-ACMW, \ce{d_{70}}-\ce{D2O}, \ce{d_{70}}-ACMW, h-\ce{D2O}, \ce{d_{13}}-\ce{D2O}, \ce{d_{13}}-ACMW. The different contrast reflectometry profiles have been offset in the \emph{y}-axis by an order of magnitude and the SLD profiles offset in the \emph{y}-axis by \SI{10e-6}{\per\square\angstrom}, for clarity.}
 \label{fig:5ns}
\end{figure}

The \SI{50}{\nano\second} production run for the Slipids potential model simulation required 13 days of using 32 cores of the SCARF computing resource.
This is non-trivial and therefore not necessarily applicable to all neutron reflectometry experiments.
However, we note that the use of a \SI{2}{\femto\second} simulation timestep could reduce this time significantly.
Additionally, Figure~\ref{fig:5ns} shows the results from the first \SI{5}{\nano\second} of the Slipid potential model, at an APM associated with a surface pressure of \SI{30}{\milli\newton\per\meter}, and already good agreement with the data is apparent.
It is important to keep in mind that this length of simulation required may be extremely system specific.
Furthermore, recent developments of molecular dynamics simulations on graphical processing units (GPUs) may allow for significant speed up of the simulations.
The nearly as accurate Berger potential model simulations (which are only marginally less accurate) took approximately 2 days, on the same compute resource.
This suggests that by using a larger timestep, shorter simulations, and the power of GPU-based molecular dynamics engines it may be possible to run these simulations alongside experiments at large facilities to aid interpretation and analysis.

\subsection{Using the Slipid simulations to improve the monolayer model}
\begin{figure}
\centering
  \includegraphics[width=0.48\textwidth]{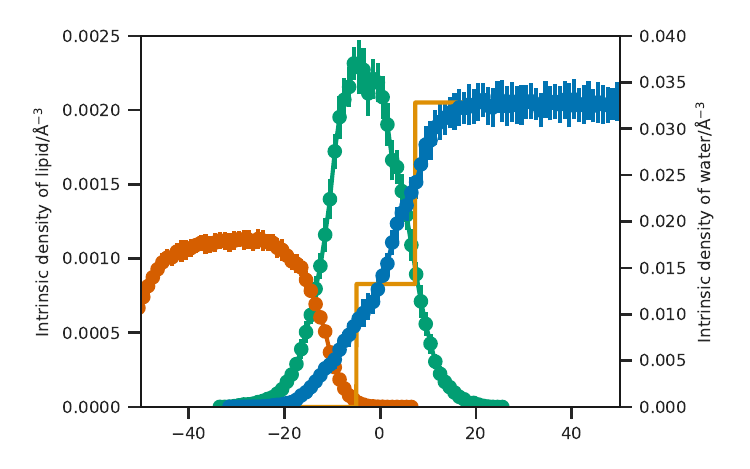}
  \caption{The simulation time-averaged intrinsic density profile of the water molecules (blue dots) and lipid components (head groups: green dots, tail groups: red dots), where the phosphorus atoms of the lipid heads create the intrinsic surface at $z=$\SI{0}{\angstrom}, at an APM associated with a surface pressure of \SI{30}{\milli\newton\per\meter} and the equivalent scattering length density from the chemically-consistent model (orange line); similar data for the other surface pressures can be found in the ESI.}
  \label{fig:waters}
\end{figure}

Despite the model layer structure offering a small improvement in agreement over the Slipid potential model simulation, we believe that it is possible to use these chemically constrained MD simulations to improve the existing monolayer model.
For example, Figure~\ref{fig:waters} considers the solvent penetration of the lipid heads, using the intrinsic surface approach to remove the effect of the interfacial roughness.
It is clear that the plot is not stepwise as is obtained from the uniform solvation model that is commonly used in traditional layer models.
Nor is the distribution sigmoidal, as there is a small deviation in the region of the ester group of the lipid heads.
This is either due to the hydrophilic interaction of the carbonyl moiety or from pockets of water forming at the air-water interface.
Regardless of the mechanism, this suggests that a different solvation model should be considered for a realistic description of the solvent penetration.

Figure~\ref{fig:waters} also shows that, without the presence of the roughness, the distribution of the head groups is relatively normal.
This agrees well with the method used previously to fit the experimental data by Hollinshead \emph{et al.} \cite{hollinshead_effects_2009}, where Gaussian functions were used to describe the lipids head and tail groups.
However, the tail group distribution is not distributed in a Gaussian fashion, and this previous method failed to account for any roughness in the interface.

Previous work has suggested that when only a single lipid type is present, the roughness between the layers should be conformal in nature, that is it should be carried uniformly through the layers \cite{kozhevnikov_general_2012,campbell_structure_2018}.
However, from the investigation of the SLD profiles in Figure~\ref{fig:ref}(b) it appears that the roughness between the lipid tails and the air is dramatically different from that at the lipid head-water interface.
In an effort to quantify the interfacial roughness in the simulations, we have used the method outlined in Section~\ref{sec:traj}.
The values for the mean, \SI{95}{\percent} quantile, and the spread between these for the \emph{z}-dimension position for atoms representative of the start, middle, and end of each of the lipid head and tails are given in Table~\ref{tab:spread}, for an APM associated with a surface pressure of \SI{30}{\milli\newton\per\meter} with the other surface pressures available in the ESI.
From this table, it is clear that at the very start of the lipid molecule (at the head) the roughness is very large with a value of \SI{\sim10}{\angstrom} for the nitrogen atom.
However this decreases slightly within the lipid head, reaching a value of 8.6\si{\angstrom} for the end of the head group.
There is then a substantial decrease noted in the lipid tail, going from \SI{\sim8.5}{\angstrom} at the start of the tail to \SI{\sim1.5}{\angstrom} at the end.
We believe that this indicates the presence of a highly non-conformal roughness in the lipid monolayer of a single lipid type and therefore in future, it is important to consider this possibility in the use of model layer structure method.
\begin{table}[h]
\small
  \caption{\ The mean, \SI{95}{\percent} quantile, and their spread for the \emph{z}-dimension position of atoms representative of difference parts of the lipid, at an APM associated with a surface pressure of \SI{30}{\milli\newton\per\meter}.}
  \label{tab:spread}
  \begin{tabular*}{0.48\textwidth}{@{\extracolsep{\fill}}llll}
    \hline
    Position & Mean/\si{\angstrom} & \SI{95}{\percent} quantile/\si{\angstrom} & Spread/\si{\angstrom} \\
    \hline
    Start-Head & 66.6 & 76.6 & 10.1 \\
    Mid-Head & 67.7 & 76.6 & 9.0 \\
    End-Head & 70.8 & 79.3 & 8.6 \\
    \hline
    Start-Tail 1 & 72.2 & 80.3 & 8.1 \\
    Start-Tail 2 & 73.0 & 81.7 & 8.6 \\
    Mid-Tail 1 & 80.9 & 87.1 & 6.2 \\
    Mid-Tail 2 & 82.3 & 87.9 & 5.6 \\
    End-Tail 1 & 91.1 & 93.3 & 2.2 \\
    End-Tail 2 & 92.4 & 93.5 & 1.1 \\
    \hline
  \end{tabular*}
\end{table}

\section{Conclusions}
This work presents, for the first time, a direct comparison between a traditional method for analysis of neutron reflectometry measurement with analysis derived from a range of all-atom and coarse-grained molecular dynamics simulations; using the all-atom Slipid, the united-atom Berger, and the coarse-grained MARTINI potential models.
It was found that the MARTINI potential model did not accurately model the lipid monolayer system, likely, due to the limitations of the 4-to-1 beading system when applied to a carbon tail containing 18 atoms.

The Berger and Slipid potential models both showed good agreement with the experimental data, however, the best agreement was obtained by the traditional monolayer model.
This would be expected given that the monolayer model contains many more ``degrees of freedom'' than the simulations which are severely chemically constrained by the potential model.

Finally, some points from the highest resolution, Slipid, simulations were noted that may be used to improve the traditional monolayer model.
For example, it is desirable to model non-uniform solvation of the head group region which would enable a more accurate modelling of the lipid monolayer and the use of a conformal roughness may not be the best constraint to apply.

\section{Author Contributions}
The initial experiments were conducted by D.J.B. and M.J.L.; the analysis methodology was developed by A.R.M. with input from J.G., A.J.S., J.L.R., S.C.P. and K.J.E.; A.R.M. wrote the manuscript, with input from all authors.

\acknowledgements{
ARM is grateful to the University of Bath and Diamond Light Source for
co-funding a studentship (Studentship Number STU0149).
This work benefited from the computing resources provided by STFC
Scientific Computing Department's SCARF cluster.
We thank Robert D. Barker for insightful discussion.}

\bibliography{paper.bib}

\end{document}